\documentclass[12pt]{article}
\usepackage{epsf}
\def\be{\begin{equation}}
\def\ee{\end{equation}}
\def\bea{\begin{eqnarray}}
\def\eea{\end{eqnarray}}

\begin{document}
\begin{titlepage}
\begin{center}
{\Large \bf William I. Fine Theoretical Physics Institute \\
University of Minnesota \\}
\end{center}
\vspace{0.2in}
\begin{flushright}
FTPI-MINN-16/22 \\
UMN-TH-3534/16 \\
July 2016 \\
\end{flushright}
\vspace{0.3in}
\begin{center}
{\Large \bf Constituent counting rule for exclusive production of heavy quarkoniumlike exotic resonance and a light hadron.
\\}
\vspace{0.2in}
{\bf  M.B. Voloshin  \\ }
William I. Fine Theoretical Physics Institute, University of
Minnesota,\\ Minneapolis, MN 55455, USA \\
School of Physics and Astronomy, University of Minnesota, Minneapolis, MN 55455, USA \\ and \\
Institute of Theoretical and Experimental Physics, Moscow, 117218, Russia
\\[0.2in]

\end{center}

\vspace{0.2in}

\begin{abstract}
The exclusive processes are considered, where a point-like source of heavy quark-antiquark pairs $Q \bar Q$, e.g. their electromagnetic current, produces  a pair consisting of a heavy quarkoniumlike exotic meson (tetraquark) or baryon (pentaquark) and a light meson or an antibaryon. For a sufficiently large mass of the heavy quark $m_Q$ there is a range of the energy $E$ above the $Q \bar Q$ threshold, where $E \ll m_Q$ and still the energy is large compared to the strong interaction scale, $E \gg \Lambda_{QCD}$. It is shown that in this energy range, where the heavy quarks are nonrelelativistic, a specific `intermediate asymptotic' behavior sets in determined by the number $n$ of the pairs of constituent quarks, with the rate scaling as $E^{1-n}$.

\end{abstract}
\end{titlepage}

Studies of the asymptotic high energy behavior of amplitudes of exclusive hadronic processes go back to the early days of the development of QCD~\cite{mmt,bf}. In particular, it has then been derived that in the ultrarelativistic regime, i.e. at the energy that is larger than any hadron masses and the scale of the strong interaction, the power of the energy in the scaling law for the fall-off the amplitudes for such processes is determined by the minimal number of constitutuent quarks in the hadrons involved in such processes. This understanding proved to be of a great practical value in numerous studies, e.g. in constructing models of hadronic form factors and in the studies of analytic properties of the amplitudes. Lately there has been a revival of interest to application of the same ideas to processes with the recently found manifestly exotic hadrons containing a heavy quark-antiquark pair in addition to light constituents, such as the isovector mesonic resonances $Z^\pm_b(10610)$ and $Z^\pm_b(10650)$~\cite{bellez} in the bottomonium sector, the charmoniumlike charged states  $\psi^\pm(4430)$~\cite{belle443,lhcb443}, $Z^\pm_c(3900)$~\cite{besz1}, $Z^\pm_c(4020)$~\cite{besz2}, and the hidden-charm pentaquarks $P_c$~\cite{lhcbp}. In particular it has been argued on the basis of the constituent counting rules~\cite{bll,bl,cks} that studies of the kinematic behavior of processes with exotic hadrons can resolve between theoretical models of their internal dynamics. These arguments however were critically analyzed in a recent paper~\cite{gmw}. 

The most basic exclusive process of a practical interest involving a heavy exotic resonance is the hard production of a pair consisting of the exotic hadron and an ordinary light meson or baryon. Particularly, the production of such pairs in $e^+e^-$ annihilation, $e^+e^- \to Z_Q \, \pi$ or $e^+e^- \to P_Q \bar p$ with $P_Q$ standing for a heavy pentaquark and $\bar p$  is the antiproton, is potentially observable in experiments at electron-positron colliders, and in fact has been observed with the mesonic resonances $Z_c$ and $Z_b$. Furthermore, the constituent counting rule in its original form~\cite{mmt,bf} was applied~\cite{kv,mv16} to description of the analytic properties of the production amplitudes. 

Clearly, the scaling behavior, based on neglecting masses of all hadrons in an exclusive process, becomes applicable only at very high energies if that process involves a heavy quark-antiquark pair. In particular, at asymptotically high energies  a heavy hidden-flavor quark pair cannot be counted as `constituent', as is pointed out in Ref.~\cite{gmw}, since in the leading order in the energy scale its production by gluons carries no suppression in comparison with light quark-antiquark pairs. It is clear however that, although formally correct, this conclusion appears to be only of an academic as opposed to practical interest. Indeed, the production amplitude falls off with the energy and becomes extremely small in the asymptotic region where the ultrarelativistic behavior for heavy quarks sets in. At a `moderate' excitation energy $E$ above the $Q \bar Q$ threshold, $\sqrt{s}=2m_Q+E$, where the amplitudes are possibly measurable in practice and where the creation and annihilation of heavy quark pairs is not essential, the behavior of the amplitudes is determined by relation between $E$ and a hadronic momentum scale $\mu$  that determines the dynamics inside the exotic states and inside ordinary light hadrons~\footnote{The effective `quenching' of the heavy quark pairs in the intermediate range of $E$ can be readily effected, for the purpose of theoretical discussion, by considering the quark and antiquark in the pair as being of different flavor.}. In light hadrons the scale $\mu$ is of order $\Lambda_{QCD}$, while in the hadrons containing a heavy $Q \bar Q$ pair this scale depends on the QCD parameters and the mass $m_Q$. In particular it becomes proportional to $\alpha_s \, m_Q$ in the limit of asymptotically heavy quark. In exotic hadrons with hidden flavors the characteristic momenta can be a mixture of low scales, that can go to very low values in loosely bound molecular states. Any detailed discussion of the internal structure of exotic heavy resonances is beyond the scope of this paper, and the notation $\mu$ is used here for a combination of those low momentum scales.  It is important for the present treatment that $\mu$ is considered to be much smaller than $m_Q$, which approximation appears to be reasonably applicable for the bottomonium sector. In the limit $E \gg \mu$ the behavior of the amplitudes becomes, to an extent, tractable by the standard in QCD methods of separation of the short- and long-distance dynamics (see e.g. a discussion of factorization in a similar context in Ref.~\cite{gmw}).

The treatment is further simplified for sufficiently heavy quarks $Q$ if simultaneously with the condition of $E$ being large as compared to $\mu$, one can also require that the excitation energy is much smaller than the heavy quark mass, $E \ll m_Q$. Clearly, the range of energy where both these restrictions apply is only marginal for the charmed quarks whose mass $m_c$ is not sufficiently larger than $\mu$, but may well be of relevance for the production of bottomonium-like exotic resonances. The condition $E \ll m_Q$ allows one to treat the heavy quarks as nonrelativistic. In what follows it will be shown that under these assumptions the rate $\Gamma$ of production by a local source $(\bar Q \Gamma Q)$ of an exclusive state $X + h$ with $h$ being a light hadron and $X$ -- an exotic resonance containing the $Q \bar Q$ heavy pair as well as light (anti)quarks scales as 
\be
\Gamma \propto E^{1-n}~,
\label{gres}
\ee
 where $n$ is the number of constituent light quark-antiquark pairs in the final state $X+h$. In particular, for the production cross section in $e^+e^-$ annihilation this implies the relations
\be
{\sigma (e^+e^- \to Z_Q \, \pi) \over  \sigma (e^+e^- \to \mu^+ \mu^-)} \propto {1 \over E}~, ~~~~ {\sigma (e^+e^- \to P_Q \, \bar p) \over  \sigma (e^+e^- \to \mu^+ \mu^-)} \propto {1 \over E^2}~.
\label{res}
\ee

\begin{figure}[ht]
\begin{center}
 \leavevmode
    \epsfxsize=16cm
    \epsfbox{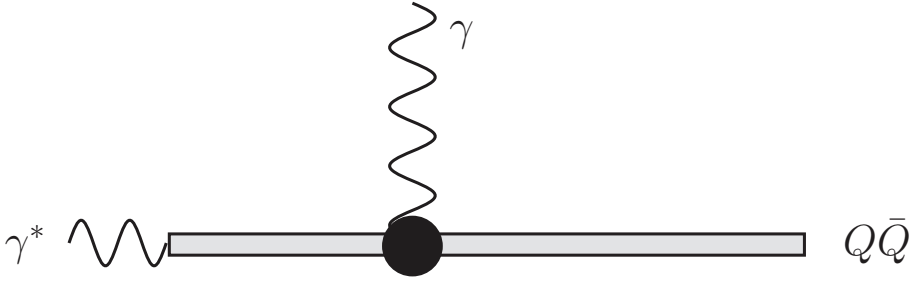}
    \caption{The graph for production by a virtual photon of $Q \bar Q$ quarkonium  with radiation of a real photon. The filled circle denotes the dipole interaction described by the Hamiltonian (\ref{hem}).   }
\end{center}
\end{figure}

The basic ingredients that lead to the scaling formula (\ref{gres}) can be illustrated starting with the simplest case $n=0$ and then increasing the number of constituent fermions. The case $n=0$ can be considered as corresponding to the process $e^+ e^- \to (Q \bar Q) + \gamma$ with the pair $(Q \bar Q)$ forming a bound (non-exotic) quarkonium state. The graph for this process is shown in Fig.~1. The propagation of the heavy quark pair is shown by a single thick box, rather than by individual lines for the quark and antiquark, reflecting the fact that for a nonrelativistic pair only the relative distance $\vec r$ between them (as a function of time) is essential. Also, due to the condition $E \ll m_Q$ the whole excess energy $E$ is carried away by the emitted photon, and any recoil of the quarkonium as whole can be neglected. Furthermore, the electromagnetic vertex for the creation of the quark pair reduces in the nonrellativistic limit to a local $\delta$-function operator ${\cal O} \to C \delta^{(3)}(\vec r)$ with the normalization constant $C$ being inessential for the present discussion of the scaling behavior. Finally, the filled circle in Fig.~1 describes the interaction of the quark pair with the electromagnetic field. For a nonrelativistic pair this interaction can be described by the Pauli Hamiltonian
\be
H_{EM}= -{2 Q  \over m_Q} \,  (\vec p \cdot \vec A)  - {Q  \over 2 m_Q} (\sigma_Q - \sigma_{\bar Q})_i B_i~,
\label{hem}
\ee
where $Q$ is the electric charge of the quark, $\vec A$ and $\vec B$ are the vector potential and the magnetic field strength for the emitted photon, $\vec \sigma_Q$ ($\vec \sigma_{\bar Q}$) are the spin operators for the quark (antiquark), and $\vec p$ stands for the momentum in the center-of-mass system. It can be noted that any spatial variation of the field of the emitted photon, set by the distance scale $\sim 1/E$, can be neglected, since the Green's function for the propagation of the pair between the local creation by the virtual photon and the emission vertex constrains the contributing distances to a much shorter scale $\sim 1/\sqrt{m_Q E}$.

The ratio of the amplitudes generated by the first term in the Hamiltonian (\ref{hem}), the electric dipole $E1$, and the second term, the magnetic dipole $M1$, is of order $p/E \sim \mu/E$. Thus at $E \gg \mu$ the dominant contribution arises from the $M1$ interaction~\footnote{This is opposite to the relation for transitions between states of a nonrelativistic bound system, where $E \sim \mu^2/m \ll m$. It can be also noted that in the discussed here process the $E1$ term describes the production of $P$-wave quarkonium, while the dominant $M1$ term corresponds to the production of $S$-wave states.}. It is important for arriving at this conclusion that it is the momentum $p \sim \mu$ that determines the emission amplitude rather than the momentum of the photon $q \approx E$, due to the gauge condition $(\vec q \cdot \vec A) = 0$. Retaining only the $M1$ term in the interaction, one readily finds that the amplitude for the process in Fig.~1 is constant in the energy:
\be
A_{\gamma} = \langle (Q \bar Q) \, \gamma| {\cal O} | 0 \rangle \propto E^0~,
\label{m1g}
\ee
since the Green's function between the vertices in Fig.~1 is of order $1/E$. The rate for the considered process is then evaluated as
\be
\sigma[e^+e^- \to (Q \bar Q) \, \gamma] \propto \int \, |A_\gamma|^2 \, 2 \pi \, \delta(E- q_0) {d^3 q \over (2 \pi)^3 \, 2 q_0} \propto E~,
\label{gg}
\ee  
which estimate agrees with Eq.(\ref{gres}) at $n=0$.

A somewhat more complex, but still  simplified example, corresponding to $n=1$ is the rather artificial process shown in Fig.~2. In this process the vector particle emitted by the heavy quark pair is virtual and produces a pair of light fermions, of which one (the fermion for definiteness) forms an `exotic' bound state $X_f$ with the $Q \bar Q$ pair and the other (antifermion) is emitted as a free particle. Since this example, discussed here purely as an illustration, is not realistic in either QED or QCD the notation `vector' (i.e. neither a photon nor gluon) and `fermion' (i.e. neither a lepton nor quark) is used. Noting that the fermion in the bound state has momentum of order $\mu$, while the antifermion carries the energy $E$, one can conclude that for the vector propagator $q^2 \sim E \mu$. Taking into account the spinor normalization factor $\sqrt{E}$ for the fast antifermion, it can be readily seen that, as far as the scaling with $E$ is concerned, the amplitude for the process in Fig.~2 contains an extra factor proportional to $1/\sqrt{E}$ in comparison with that for a real photon emission in Fig.~1. Thus the rate for the (unrealistic) process $e^+e^- \to X_f \, \bar f$ scales as $1/E$ also in agreement with Eq.(\ref{gres}).

\begin{figure}[ht]
\begin{center}
 \leavevmode
    \epsfxsize=16cm
    \epsfbox{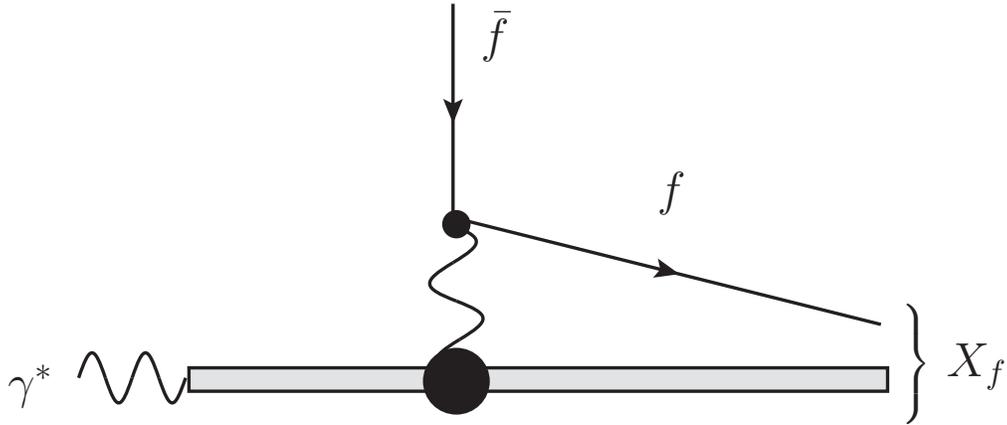}
    \caption{An (artificial) illustrative process with creation of a pair of light fermions $f \bar f$ with $f$ forming an exotic bound state with the heavy $Q \bar Q$ pair and the antifermion $\bar f$ emitted with energy $E$.    }
\end{center}
\end{figure}

The simplest process involving production of an actual exotic quarkoniumlike resonance and a light meson is $e^+ e^- \to Z_Q \pi$. This is the process that, for concreteness, is discussed here, since the treatment is trivially generalized to any similar production of a heavy exotic four-quark resonance in association with a light meson. The graphs with hard production of light quark pairs in this process are shown in Fig.~3. The relevant terms in the interaction of  a nonrelativistic heavy quark pair with gluons are described by the Hamiltonian
\be
H_{QCD}=  -{t^a_Q - t^a_{\bar Q}  \over m_Q} \,  (\vec p \cdot \vec A^a)  - {t^a_Q - t^a_{\bar Q}   \over 4 m_Q} (\sigma_Q - \sigma_{\bar Q})_i B^a_i +  T^a A^a_0~, 
\label{hqcd}
\ee
where $A^a$ and $B^a$ are the potential and the magnetic strength of the gluon field, $t^a_Q$ ($t^a_{\bar Q}$) are the color generators for the heavy quark (antiquark) and $T^a = t^a_Q + t^a_{\bar Q}$ is the total color generator for the $Q \bar Q$ system. 

\begin{figure}[ht]
\begin{center}
 \leavevmode
    \epsfxsize=16cm
    \epsfbox{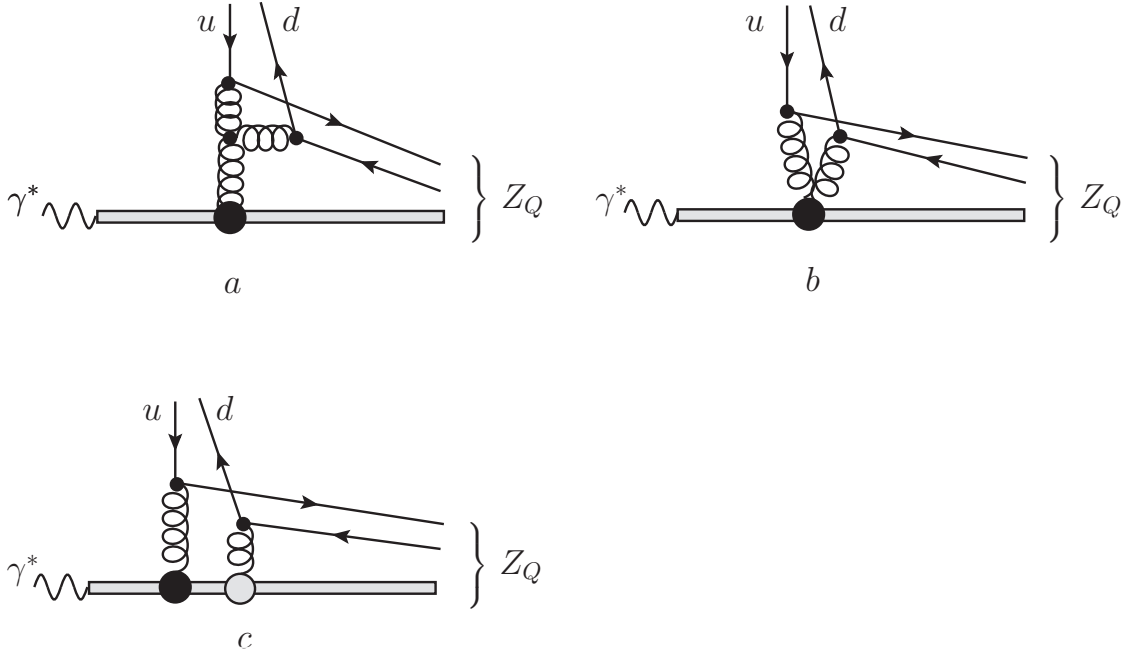}
    \caption{The mechanisms for exclusive production of the pairs $Z_Q \pi$. The large black filled circle denotes the color dipole interaction given by the two first terms in the Hamiltonian (\ref{hqcd}), while the greyed circle corresponds to the color monopole described by the last term. }
\end{center}

\end{figure}

The last term in the Hamiltonian (\ref{hqcd}) is the monopole term. Unlike the first two terms, its contribution is not suppressed by the heavy quark mass and it would be dominant for a color octet pair. However, the source (the electromagnetic current) produces a color singlet $Q \bar Q$ pair, hence the first emission of a hard gluon is possible only due to the first two terms. These terms contain the operator $t^a_Q - t^a_{\bar Q}$ converting the pair to color octet state, so that in the subsequent emissions from the heavy system only the monopole term can be retained in the leading order in $m_Q$. (In particular, the chromomagnetic term proportional to $T^a$, not shown in Eq.(\ref{hqcd}), is totally negligible because of its suppression by $1/m_Q$.)  Finally, the graph in Fig.~3$b$ arises from the quadratic in $A^a$ term in the chromomagnetic field $B^a$. 

It can be further noticed that the large components, proportional to $E$, of the momenta of the fast light quarks as well as of the virtual gluons are collinear and proportional to the momentum of the emitted pion. For this reason the virtuality of each of the gluon propagators is $q^2 \sim E \mu$. Another consequence of the collinearity of the large components is that, as in the previously discussed simplified cases, due to the gauge condition the contribution of the chromoelectric $E1$ term from Eq.(\ref{hqcd}) does not contain a large momentum proportional to $E$ and is thus suppressed relative to that of the $M1$ chromomagnetic dipole.

One can readily find that the contribution of the graphs of Figs.~3$a$ and 3$b$ to the amplitude is of the same order in the energy $E$, and including the spinor normalization factors, proportional to $\sqrt{E}$ for each fast (anti)quark, the energy dependence of this contribution can be evaluated as
\be
A(Z_Q \pi) \sim {1 \over E}~,
\label{azpi}
\ee
where it is also taken into account that the three gluon vertex in Fig.~3$a$ is proportional to $E$. On the other hand, the contribution from graph in Fig.~3$c$ is only of the order $1/E^2$ and is thus subdominant. This is because the extra propagator of the heavy pair introduces the factor $1/E$ with no energy dependence of the monopole vertex, while an extra hard gluon propagator and the three gluon vertex (in the graph of Fig.~3$a$) result in the factor $E/(E \mu) = 1/\mu$. It can also be readily verified that the graphs where the additional light quark pair is emitted by a gluon attached to another light quark line are suppressed relative to (\ref{azpi}) by a factor $1/\sqrt{E}$, and for this reason are not considered in the present discussion.

The $E$ dependence of the rate  generated by the amplitude (\ref{azpi}) can be estimated as
\be
\sigma(e^+e^- \to Z_Q \pi) \propto \int \, |A(Z_Q \pi)|^2 \, \delta(E-\omega_1 - \omega_2) \, {d^3 k_1 \, d^3 k_2 \over \omega_1 \, \omega_2} \propto {1 \over E}~,
\label{n2}
\ee
where $k_1$ and $k_2$ ($\omega_1$ and $\omega_2$) are the momenta (energies) of the fast quark and antiquark. The large longitudinal components of the momenta cancel against the energies in the denominator, while the integration over the relative transverse momentum is constrained at $\mu^2$ by the condition that the light quark and antiquark make a pion. The only large factor remaining in the integration arises from the integration over the total  momentum of the pion, and, together with the energy conservation $\delta$ function gives a factor of $E$, i.e. the same as in the previously considered cases $n=0$, and $n=1$. The final estimate of the energy dependence in Eq.(\ref{n2}) is obviously the one given by the general formula (\ref{gres}). 

The generalization of the derivation of Eq.(\ref{gres}) to the case of arbitrary $n$ is quite straightforward. Indeed, as argued for the case of $n=2$, the dominant $E$ dependence arises from a single hard $M1$ interaction on the line of the heavy pair, while graphs with any additional vertices on this line produce only a subdominant contribution. Thus emission of additional constituent light quark pairs proceeds through the branching of the gluons in the graphs of Figs.~3$a$ and 3$b$. Each such branching gives in the amplitude an extra factor proportional to $1/\sqrt{E}$. On the other hand, the phase space integration does not introduce new energy dependence once the condition that $n$ produced fast (anti)quarks are constituents in a fast hadron. Thus one concludes that each extra pair of produced constituent light quarks brings the factor $1/E$ in the rate, and thus arrives at the general formula (\ref{gres}).

Before concluding, two points related to the derived here scaling rule and the mechanism leading to the derivation merit a brief discussion. One point is regarding the spin state of the heavy quark pair corresponding to the dominant at large $E$ production mechanism in $e^+e^-$ annihilation. Namely, the electromagnetic current produces the $Q \bar Q$ pair in the spin triplet state. The spin operator $(\vec \sigma_Q - \vec \sigma_{\bar Q})$ flips the total spin into the siglet state. Thus the dominance of the $M1$ interaction in the considered energy region implies that in the exclusive production of  the heavy exotic resonances in a pair with a light hadron there should be mostly the states with a spin singlet heavy quark pair. It is not clear at present whether this behavior can be studied in experiments. Indeed, the only so far known bottomoniumlike exotic resonances  $Z_b(10610)$ and $Z_b(1065)$ are mixed states with regards to the total spin of the $b \bar b$ pair~\cite{bgmmv}, and can thus be produced through the spin singlet component. It would however be possible to study the predicted behavior if some or all of the expected~\cite{mvwb} isovector $G$-negative bottomoniumlike resonances $W_{bJ}$ are found and become accessible to observation in $e^+e^-$ annihilation through $e^+e^- \to W_{bJ} \rho$. Two of these resonances with $J=0$: $W_{b0}$ and $W'_{b0}$, also contain a spin singlet heavy quark component and thus their exclusive production at energy well above the threshold, should have a higher yield than for the resonances $W_{b1}$ and $W_{b2}$ containing only pure spin triplet $b \bar b$ quark pair. 

Another point that merits mentioning is that the rather slow $1/E$ fall off of the cross section for $e^+e^- \to Z_b \, \pi$ generally implies that there should be some production of this exclusive final state in the continuum at energies above the region of the $\Upsilon(nS)$ resonances. At present it does not appear possible to reliably estimate the rate beyond the simple remark that it contains an extra suppression by the inverse of the mass $m_b$ inherent in the $M1$ interaction in Eq.(\ref{hqcd}).  Namely, the relations (\ref{res}) with proper dimensional parameters restored should read as
\be
{\sigma (e^+e^- \to Z_Q \, \pi) \over  \sigma (e^+e^- \to \mu^+ \mu^-)} \sim {\mu^3 \over m_Q^2 \, E}~, ~~~~ {\sigma (e^+e^- \to P_Q \, \bar p) \over  \sigma (e^+e^- \to \mu^+ \mu^-)} \sim {\mu^4 \over m_Q^2 \, E^2}~.
\label{resd}
\ee
As of yet the production of the final states $Z_b \, \pi$ has been observed~\cite{belle1508} only in the $\Upsilon(5S)$ and $\Upsilon(6S)$ resonances.  It would thus be quite interesting if a nonresonant production of the $Z_b \, \pi$ pairs could be studied experimentally at energies above the $\Upsilon(6S)$ resonance.

This work is supported in part by U.S. Department of Energy Grant No.\ DE-SC0011842.


\begin{thebibliography}{99}

\bibitem{mmt} 
  V.~A.~Matveev, R.~M.~Muradian and A.~N.~Tavkhelidze,
  Lett.\ Nuovo Cim.\  {\bf 7}, 719 (1973).
\bibitem{bf} 
  S.~J.~Brodsky and G.~R.~Farrar,
  Phys.\ Rev.\ D {\bf 11}, 1309 (1975).
	
\bibitem{bellez} 
  A.~Bondar {\it et al.}  [Belle Collaboration],
  Phys.\ Rev.\ Lett.\  {\bf 108}, 122001 (2012)
  [arXiv:1110.2251 [hep-ex]].
\bibitem{belle443} 
  S.~K.~Choi {\it et al.} [Belle Collaboration],
  Phys.\ Rev.\ Lett.\  {\bf 100}, 142001 (2008)
  doi:10.1103/PhysRevLett.100.142001
  [arXiv:0708.1790 [hep-ex]].
\bibitem{lhcb443} 
  R.~Aaij {\it et al.} [LHCb Collaboration],
  Phys.\ Rev.\ Lett.\  {\bf 112}, no. 22, 222002 (2014)
  doi:10.1103/PhysRevLett.112.222002
  [arXiv:1404.1903 [hep-ex]].
	
\bibitem{besz1} 
  M.~Ablikim {\it et al.} [BESIII Collaboration],
  Phys.\ Rev.\ Lett.\  {\bf 110}, 252001 (2013)
  doi:10.1103/PhysRevLett.110.252001
  [arXiv:1303.5949 [hep-ex]].	
	
\bibitem{besz2} 
  M.~Ablikim {\it et al.} [BESIII Collaboration],
  Phys.\ Rev.\ Lett.\  {\bf 111}, no. 24, 242001 (2013)
  doi:10.1103/PhysRevLett.111.242001
  [arXiv:1309.1896 [hep-ex]].
	
\bibitem{lhcbp} 
  R.~Aaij {\it et al.} [LHCb Collaboration],
  Phys.\ Rev.\ Lett.\  {\bf 115}, 072001 (2015)
  doi:10.1103/PhysRevLett.115.072001
  [arXiv:1507.03414 [hep-ex]].
\bibitem{bll} 
  S.~H.~Blitz and R.~F.~Lebed,
  Phys.\ Rev.\ D {\bf 91}, no. 9, 094025 (2015)
  doi:10.1103/PhysRevD.91.094025
  [arXiv:1503.04802 [hep-ph]].
		
\bibitem{bl} 
  S.~J.~Brodsky and R.~F.~Lebed,
  Phys.\ Rev.\ D {\bf 91}, 114025 (2015)
  doi:10.1103/PhysRevD.91.114025
  [arXiv:1505.00803 [hep-ph]].
	
\bibitem{cks} 
  W.~C.~Chang, S.~Kumano and T.~Sekihara,
  Phys.\ Rev.\ D {\bf 93}, no. 3, 034006 (2016)
  doi:10.1103/PhysRevD.93.034006
  [arXiv:1512.06647 [hep-ph]].
	
\bibitem{gmw} 
  F.~K.~Guo, U.~G.~Mei{\ss}ner and W.~Wang,
  arXiv:1607.04020 [hep-ph].
	
\bibitem{kv} 
  V.~Kubarovsky and M.~B.~Voloshin,
  Phys.\ Rev.\ D {\bf 92}, no. 3, 031502 (2015)
  doi:10.1103/PhysRevD.92.031502
  [arXiv:1508.00888 [hep-ph]].
\bibitem{mv16} 
  M.~B.~Voloshin,
  Phys.\ Rev.\ D {\bf 94}, no. 1, 014004 (2016)
  doi:10.1103/PhysRevD.94.014004
  [arXiv:1604.08196 [hep-ph]].
	
\bibitem{bgmmv}
  A.~E.~Bondar, A.~Garmash, A.~I.~Milstein, R.~Mizuk, M.~B.~Voloshin,
  Phys.\ Rev.\  {\bf D84}, 054010 (2011).
  [arXiv:1105.4473 [hep-ph]].
	
\bibitem{mvwb} 
  M.~B.~Voloshin,
  Phys.\ Rev.\ D {\bf 84}, 031502 (2011)
  doi:10.1103/PhysRevD.84.031502
  [arXiv:1105.5829 [hep-ph]].
		
\bibitem{belle1508} 
  A.~Abdesselam {\it et al.} [Belle Collaboration],
  arXiv:1508.06562 [hep-ex].
	
	

\end{thebibliography}
\end{document}